\documentstyle{elsart}
\input{psfig}
\begin{document}

\begin{frontmatter}

\hfill INP 1801/PH

\vspace{0mm}

\title{Solitons in a chiral quark model with non-local
interactions\thanksref{grants}}
\thanks[grants]{Research supported by
        the Polish State Committee for
Scientific Research grant 2P03B-080-12, and by the
Ministry of Science of Slovenia,
Scientific and Techological Cooperation Joint Project between
Poland and Slovenia}
\thanks[emails]{%
\hspace{0mm} E-mail: Bojan.Golli@ijs.si,
broniows@solaris.ifj.edu.pl,
ripka@spht.saclay.cea.fr}
\author[LJ]{Bojan Golli},
\author[INP]{%
Wojciech Broniowski}, and
\author[Saclay]{Georges Ripka}
\address[LJ]
{Faculty of Education, University of Ljubljana and J.~Stefan Institute,
Jamova 39, P.O.~Box 3000, 1001 Ljubljana, Slovenia}
\address[INP]{H. Niewodnicza\'nski Institute of Nuclear
Physics,
         PL-31342 Krak\'ow, Poland}
\address[Saclay] {Service de
Physique Th\'eorique, Centre d'Etudes de Saclay,
F-91191 Gif-sur-Yvette Cedex, France}

\begin{abstract}
Hedgehog solitons are found in a chiral quark model with non-local
interactions. The solitons are stable without the
chiral-circle constraint for the
meson fields, as was assumed in previous Nambu-Jona--Lasinio model
with local interactions.
\end{abstract}

\begin{keyword} Effective chiral models, chiral solitons, models of baryons
\end{keyword}

\end{frontmatter}


\vspace{-7mm}

\noindent
PACS: 12.39.-x, 12.39.Fe, 12.40.Yx

In recent years a lot of efforts have been undertaken to describe baryons as
solitons of effective chiral models. Of particular interest are models which
include the Dirac sea, since they allow for a unified description of mesons
and baryons. So far solitons have only been obtained in the
Nambu--Jona-Lasinio model with the proper-time or Pauli-Villars
regularization of the Dirac sea \cite{Goeke96,Reinhardt96,Ripka97}.\footnote{%
Recently solitons have also been reported in the Nambu--Jona-Lasinio model
with the sharp 3-momentum cut-off, but only in the Thomas-Fermi
approximation for the Dirac sea \cite{Coim97}. These solitons cease
to exist when the mean-field calculation of the Dirac sea is carried out
exactly.} One problem encountered with the proper-time regularization, by far
the most commonly used, is that solitons turn out to be
unstable unless the sigma and pion fields are constrained to the chiral
circle \cite{Goeke92,Ripka93d}. Such a constraint is external to the model
defined at the quark level. Furthermore, somewhat artificially, the
regularization in these models is applied only to the real (non-anomalous)
part of the Euclidean quark loop term. The (finite) imaginary part is left
unregularized in order to properly describe anomalous processes. The
above-mentioned problems disappear in models with non-local quark
interactions, which result, for example, from QCD derivations of low-energy
effective theories \cite{Diakonov86,Ball90,Roberts92}. In such models, the
quark propagators have 4-momentum-dependent masses which regularize the
theory in the Euclidean domain. This regularization is in fact far more
satisfying than in local models. Indeed, it is no longer required to
regularize differently the real and imaginary parts. Furthermore, anomalous
processes turn out not to depend on the cut-off, and are properly
described \cite{Ripka93,Cahill87,Holdom89}.
Finally, the momentum-dependent regulator makes
the theory finite to all orders in the $1/N_c$ expansion. This is in
contrast to local models, where inclusion of higher-order-loop effects
requires extra regulators \cite{Ripka96b}, and the predictive power is
thereby reduced. The price to pay is a greater difficulty in calculating
hadron properties. In view of the afore-mentioned advantages, it is a
challenge to discover whether solitons exist in these theories. The purpose
of this Letter is to report that indeed stable solitons do exist in a chiral
quark model with non-local interactions.

Our model is defined by the Euclidean action:
\begin{equation}
I=-{\rm Tr}\log \left( -i\partial _\mu \gamma _\mu +m+r\Phi r\right) +\frac
1{2G^2}\int d_4x\left( S^2+P_a^2\right)   \label{action}
\end{equation}
where $\Phi =S+i\gamma _5P_a\tau _a$ is the $SU(2)$ chiral field. The chiral
field is the dynamical variable of the system and it is local in coordinate
space. The trace is in color, isospin, Dirac indices and space-time. The
average $u$ and $d$ current quark mass is denoted by $m$. The coupling
constant $G$ has the dimension of inverse energy. The operator $r$ is the
regulator, chosen to be a scalar function diagonal in the 4-momentum space:
\mbox{$\langle k|r|k^{\prime }\rangle
=\delta (k-k^{\prime })r_k$}. It introduces a non-locality in the interaction
between quarks and the chiral field. The particular way of introducing the
non-locality is the one which results from instanton-induced interactions
\cite{Diakonov86}. It has also been used in Ref.~\cite{Krewald92}.\footnote{%
With help of the semi-bosonization technique, it is straightforward to show
that the model (\ref{action}) is equivalent to the model of Birse {\em et al.%
} \cite{PlantB,BowlerB}, defined by introducing a non-local separable 4-quark
interaction of the form \mbox{$%
\delta(p_1+p_2+p_3+p_4)r_{p_1}r_{p_2}r_{p_3}r_{p_4}$}, where $p_i$ are the
(incoming) momenta of quarks. The meson fields of Eq.~(\ref{action}) are
related to the quark fields of Refs.~\cite{PlantB,BowlerB} in the following
way:
\[
\left\{ S(x),P_a(x)\right\} =-G^2\int \frac{d^4p_1}{(2\pi )^4}\int \frac{%
d^4p_2}{(2\pi )^4}\;\e^{-i(p_1+p_2)\cdot x}\overline{\psi }%
(p_1)r_{p_1}\left\{ 1,i\gamma _5\tau _a\right\} r_{p_2}\psi (p_2).
\]
} We choose the Gaussian form $r_k=e^{-\frac{k^2}{2\Lambda ^2}}$, where $%
\Lambda $ is a cut-off parameter. The form of $r$ is not critical for
results presented here.

Our treatment of model (\ref{action}) is the same as in other soliton
calculations in effective chiral models: we treat the chiral $S$ and $P_a%
$ fields classically, which amounts to keeping the leading-order
contribution in the number of colors, $N_c$. For sufficiently large values of
$G$ the model leads to dynamical chiral symmetry breaking. The vacuum value
of the scalar field, $S_0$, is determined by the stationary-point condition
\mbox{$\left. \delta I/\delta
S\right| _{S=S_{0,}P_a=0}=0$}, which explicitly gives:
\begin{equation}
\frac 1{G^2}=4N_cN_f\int \frac{d_4k}{\left( 2\pi \right) ^4}\,\frac{%
r_k^2\left( r_k^2+\frac m{S_0}\right) }{k^2+\left( m+r_k^2S_0\right) ^2}.
\label{gap}
\end{equation}
The pion mass, $m_\pi =139\,$MeV, is identified with the location of the pole
of the pion propagator, and the pion decay constant, $F_\pi =93\,$MeV, is
easily obtained via the Gell-Mann-Oakes-Renner relation \cite{PlantB,BowlerB}%
. We keep the leading terms in the chiral expansion:
\begin{equation}
F_\pi ^2=2N_cN_ff+{\cal O}\left( m^2\right) ,\quad \quad \quad m_\pi ^2=%
\frac{2mg}{S_0f}+{\cal O}\left( m^2\right) ,  \label{fpimpi}
\end{equation}
where
\begin{equation}
g=\int \frac{d_4k}{\left( 2\pi \right) ^4}\frac{r_k^2}{k^2+\left(
r_k^2S_0\right) ^2}\quad \mbox{and}\quad f=\int \frac{d_4k}{\left( 2\pi
\right) ^4}\frac{r_k^4-k^2r_k^2\frac{dr_k^2}{dk^2}+k^4\left( \frac{dr_k^2}{%
dk^2}\right) ^2}{\left( k^2+\left( r_k^2S_0\right) ^2\right) ^2}.  \nonumber
\end{equation}
The model (\ref{action}) has three parameters: $G$, $\Lambda$, and $m$. The
stationary-point condition (\ref{gap}) and the equations (\ref{fpimpi})
allow us to express them in terms of one parameter, $S_0$. Our numerical
results are presented for various choices of $S_0$.\footnote{%
It should be noted that unlike the NJL model with local interactions, $S_0$
does not have the interpretation of the constituent quark mass. It does not
correspond to the location of the pole in the quark propagator, which is
obtained by the condition $k^2+(r_k^2S_0+m)^2=0$. In fact,
in the vacuum sector, for sufficiently
large $S_0$ the quark propagator has no poles on the
physical axis, {\em i.e.} for real negative $k^2$. See the
discussion following of Fig.~%
\ref{fig2}.}

The static meson fields are determined self-consistently by solving the
Euler-Lagrange equations deduced from the action (\ref{action}). In order to
evaluate the trace in functional space 
it is convenient to introduce the energy-dependent Dirac Hamiltonian, $%
h\left( \nu ^2\right) $, given by
\begin{equation}
h\left( \nu ^2\right) =-i\vec \alpha \cdot \nabla +\beta r\left( \nu ^2-\vec
\nabla ^2\right) \Phi \,r\left( \nu ^2-\vec \nabla ^2\right) +\beta m,\qquad
\nonumber
\end{equation}
with eigenstates $\left| \lambda _\nu \right\rangle $ and eigenvalues $%
e_\lambda \left( \nu ^2\right) $ satisfying
\mbox{$h\left( \nu ^2\right) \left|
\lambda _\nu \right\rangle =e_\lambda \left( \nu ^2\right) \left| \lambda
_\nu \right\rangle$}.
We obtain
\begin{equation}
\left\{ S\left( \vec x\right) ,P_a\left( \vec x\right) \right\} =G^2\int
\frac{d\nu }{2\pi }\sum_\lambda \frac{\langle \lambda _\nu |r|\vec x\rangle
\left\{ \beta ,i\beta \gamma _5\tau _a\right\} \langle \vec x|r|\lambda _\nu
\rangle }{i\nu +e_\lambda \left( \nu ^2\right) }.  \label{eulag}
\end{equation}
The integration contour over $\nu $ in Eq. (\ref{eulag}) has to be chosen in
such a way as to include the {\em valence} quark orbit. This point requires
a careful explanation. It is well known that in hedgehog solitons a $0^{+}$
(grand-spin zero, positive-parity) orbit becomes bound and well isolated
from other orbits \cite{Ripka84}. This feature persists in the present model.
We search for a pole of the quark
propagator in the complex $\nu $ plane by solving the equation 
\begin{equation}
i\nu +e_0\left( \nu ^2\right) =0,  \label{eval}
\end{equation}
where $e_0\left( \nu ^2\right) $ is the eigenvalue of the valence $0^{+}$
orbit calculated with the Dirac Hamiltonian $h\left( \nu ^2\right) $. The
value of $\nu $ which solves Eq. (\ref{eval}) is denoted by $ie_{{\rm val}}$%
, and the corresponding orbit by $|{\rm val}\rangle $. We find that it is
always possible to find a solution of Eq.~(\ref{eval}) in the background of
soliton fields, even for choices of model parameters for which the quarks
do not materialize on-shell in the vacuum \cite{Ripka97,Krewald92,Roberts92b}%
. The solution of Eq.~(\ref{eval}) corresponds to a pole of the quark
propagator which lies on the imaginary ({\em i.e.} physical) axis close to $%
\nu =0$. It is well separated from other (unphysical) poles in the complex
plane. Such poles are associated with the presence of the regulator \cite
{Ripka95}. In all cases described in this Letter $e_{{\rm val}}>0$. The
inclusion of the valence orbit in Eq. (\ref{eulag}) yields
\begin{eqnarray}
\left\{ S\left( \vec x\right) ,P_a\left( \vec x\right) \right\} 
&=&G^2\langle {\rm val}|r|\vec x\rangle \left\{ \beta ,i\beta \gamma _5\tau
_a\right\} \langle \vec x|r|{\rm val}\rangle   \nonumber \\
&&+G^2\int_{-\infty }^\infty \frac{d\nu }{2\pi }\sum_\lambda \frac{\langle
\lambda _\nu |r|\vec x\rangle \left\{ \beta ,i\beta \gamma _5\tau _a\right\}
\langle \vec x|r|\lambda _\nu \rangle }{i\nu +e_\lambda \left( \nu ^2\right) 
}.  \label{eulag2}
\end{eqnarray}
The second term is the Dirac sea contribution, where the integral over $\nu $
is performed numerically along the real $\nu $-axis. Note that in the
absence of the regulator ($r=1$) the eigenvalues are energy-independent, the
integral over $\nu $ can be done via the Cauchy theorem, and the Dirac-sea part
of Eq. (\ref{eulag}) reduces to the usual formula expressed by the sum over
the negative-energy spectrum.

There is a non-trivial problem associated with theories with non-local
interactions: Noether currents, in particular the baryon current, acquire
extra contributions induced by the presence of the regulator $r$
\cite{Ripka93,BowlerB,PlantB}. By
performing the gauge transformation, $\psi \left( x\right) \rightarrow
e^{i\eta \left( x\right) }\psi \left( x\right) $, the quark part of the
action is transformed into: 
\begin{equation}
-{\rm Tr}\log e^{-i\eta }\left( -i\partial _\mu \gamma _\mu +r\Phi r\right)
e^{i\eta }=-{\rm Tr}\log \left( -i\partial _\mu \gamma _\mu +\frac{\partial
\eta }{\partial x_\mu }\gamma _\mu +e^{-i\eta }r\Phi re^{i\eta }\right)
\label{iq}
\end{equation}
The extra contributions are due to the term $e^{-i\eta }r\Phi re^{i\eta }$.
When these are taken into account, the expression for the baryon number
becomes: 
\begin{equation}
B=\frac 1i\frac{\delta I(\eta )}{\delta \partial \eta /\partial x_0}=-\frac
1{2\pi iN_c}\int d\nu \sum_\lambda \;\frac{i+\frac{de_\lambda \left( \nu
^2\right) }{d\nu }}{i\nu +e_\lambda \left( \nu ^2\right) }.  \label{bb4}
\end{equation}
The non-local regulator contributes the extra term 
\mbox{$\frac{d e_\lambda
\left( \nu ^2\right) }{d \nu }$}, which causes the residues of poles of the
quark propagator to be normalized to $1$.\ As a result, the baryon number is
properly quantized, independently of the shape of the regulator.\footnote{%
Note that the baryon number in not quantized (although it is enforced to be
equal to one) in the approach of Ref.~\cite{Schlienz}, where the proper-time
regularization is applied to the imaginary (anomalous) part of the effective
action.}

The energy of the time independent soliton can be deduced from the euclidean
action (\ref{action}): 
\begin{equation}
E_{{\rm sol}}=N_ce_{{\rm val}}+\frac 1{2\pi }\int_{-\infty }^\infty \nu d\nu
\sum_\lambda \frac{i+\frac{de_\lambda }{d\nu }}{i\nu +e_\lambda \left( \nu
^2\right) }+\frac 1{2G^2}\int d_3x\left( S^2+P_a^2\right) -{\rm vac},
\label{esol}
\end{equation}
where ${\rm vac}$ denotes the vacuum subtraction. The same expression can be
derived from the Noether construction of the energy-momentum tensor.

\begin{figure}[h]
\centerline{\psfig{figure=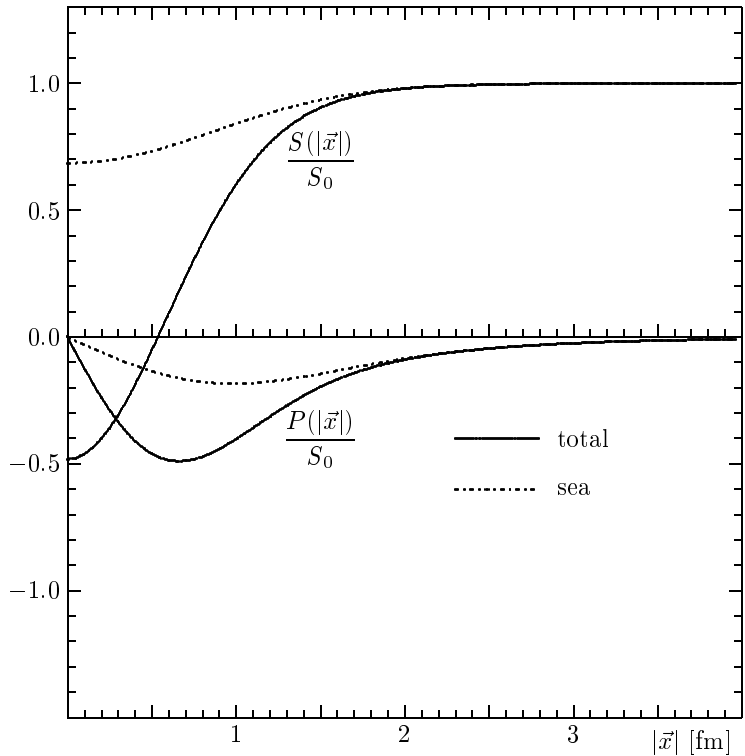,height=100mm}}
\caption{Self consistent fields for $S_0=400$~MeV}
\label{fig}
\end{figure}
Our numerical calculations were performed with the hedgehog-shaped chiral
field, in which $S(\vec x)=S(|\vec x|)$ and $P_a(\vec x)=\hat x_aP(|\vec x|)$%
. The Euler-Lagrange equations (\ref{eulag2}) are solved by iteration.\ The
Kahana-Ripka spherical plane-wave basis
\cite{Ripka84} is well suited to
calculate the matrix elements of the Dirac Hamiltonian $h\left( \nu
^2\right)$. The quark orbits are calculated by diagonalizing $h\left( \nu
^2\right)$ in this basis, successively for each value of the energy $\nu$.
We find convergence {\em without forcing the fields to remain on the chiral
circle}, as was the case of solitons in
Refs. \cite{Goeke96,Reinhardt96,Ripka97}. Figure \ref%
{fig} shows the self-consistent fields obtained for $S_0=400$~MeV. The
departure from the chiral circle is substantial. 
Increasing $S_0$ brings the fields closer to the chiral circle. It is
important to stress that our non-local regularization cuts off effects due to
high gradients in the chiral fields. Such effects were responsible for the
instability in the case of the proper-time regularization \cite
{Goeke92,Ripka93d}.

\begin{figure}[th]
\centerline{\psfig{figure=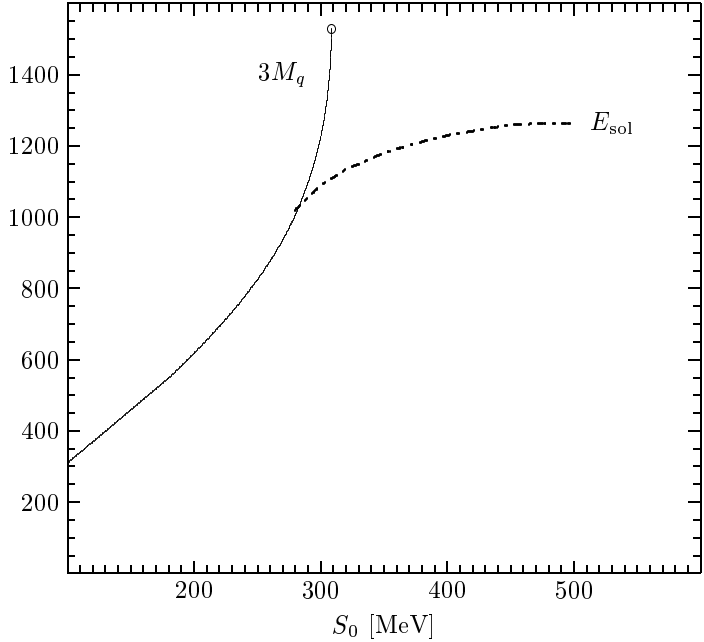,height=85mm}}
\caption{The energy of the soliton (dot-dashed line) and the three
free-space quark mass (solid line) plotted as functions of the parameter $S_0
$. Above $S_0 \simeq 280{\rm MeV}$ the soliton is energetically stable.}
\label{fig2}
\end{figure}
Figure \ref{fig2} shows that our soliton is energetically stable above $%
S_0\simeq 280{\rm MeV}$. We plot the soliton mass (dot-dashed line) and 3
times the vacuum quark mass, $3M_q$ (solid line). The quantity $M_q$ is
defined as the pole of the quark propagator in the vacuum, {\em i.e.}
corresponds to the solution of the equation $\left.
k^2+(r_k^2S_0+m)^2\right| _{k^2=-M_q^2}=0$. Real solutions to this equation
exist only below the critical point denoted by a circle at $S_0=309{\rm MeV}$%
. Thus in the region $280{\rm MeV}<S_0<309{\rm MeV}$ the mass of the soliton
is lower than the mass of three quarks, hence the soliton is bound. Above $%
S_0=309{\rm MeV}$ the soliton is also stable, since there are no asymptotic
quark states (no real quark poles) into which it could decay to. Below $%
S_0=280{\rm MeV}$ the soliton is not energetically stable.

Table \ref{table} lists several soliton properties with various values of
the vacuum scalar field. Parameters $\Lambda $ and $m$ are determined from (%
\ref{fpimpi}), and $G$ from the stationarity point condition (\ref{gap}).
Different contributions to the soliton energy are given: the energy of the
valence orbit obtained from (\ref{eval}), the Dirac sea contribution (the
second term in (\ref{esol})). The contribution of the third term in (\ref
{esol}) is not explicitly given since it can be deduced from the total
energy and other components given in Table \ref{table}. 
\begin{table}[tbp]
\centering
\begin{tabular}{|c|c|c|c||c|c|c|c|c|}
\hline
$S_0$ & $\Lambda$ & $m$ & $1/G$ & $e_{{\rm val}}$ & $E_{{\rm Dirac}}$ & $E_{%
{\rm sol}}$ & $\langle r^2\rangle^{1/2}$ & $g_A$ \\ 
MeV & MeV & MeV & MeV & MeV & MeV & MeV & fm &  \\ \hline
300 & 760 & 7.62 & 182 & 295 & 2360 & 1088 & 1.32 & 1.28 \\ 
350 & 627 & 10.4 & 140 & 280 & 1713 & 1180 & 1.04 & 1.16 \\
400 & 543 & 13.2 & 113 & 272 & 1433 & 1229 & 0.97 & 1.14 \\ 
450 & 484 & 15.9 & 94 & 266 & 1273 & 1260 & 0.96 & 1.12 \\ \hline
\end{tabular}
\caption{Properties of the self-consistent soliton solutions}
\label{table}
\end{table}
The soliton mean square baryon radius $\langle r^2\rangle $ is calculated
from the baryon density. Its main contribution is due to the valence quarks,
and the Dirac sea contributes only about 2~\%. The axial-vector charge of
the nucleon, $g_A$, is deduced from the asymptotic behavior of the pion
field, given by \mbox{$P(r)\sim -D(1/r^2+m_\pi /r){\rm exp}(-m_\pi r)$},
and equals
to \cite{ANW83,CB86} \mbox{$g_A=8\pi DF_\pi /3$}.
For $S_0$ above 300 MeV the value
of $g_A$ is somewhat smaller than the experimental value of 1.26, but 50~\%
larger than the leading-$N_c$ value obtained using proper time
regularization and constraining the chiral field to remain on the chiral
circle \cite{Goeke96}. Our value of $g_A$ is intermediate between too low
Skyrme values (on the chiral circle) and too high values obtained in
T.~D.~Lee solitons (where the pion field vanishes) which are more akin to
the MIT bag model.

Our soliton is a mean-field hedgehog solution which, of course, should not
be identified with the nucleon. Quantizing the collective coordinates, or
projecting the soliton wavefunction on the subspace with nucleon quantum
numbers, will reduce its energy by eliminating the spurious rotational and
translational motion. It is therefore not undesirable that the energies in
Table \ref{table} be higher than the experimental nucleon mass. We may
similarly expect that the large radius of the soliton will be reduced when
the center-of-mass corrections are calculated. The aim in this Letter was to
investigate the mechanism of soliton formation and its stability in chiral
models with non-local regulators, rather than to make detailed predictions
for various nucleon observables. Further properties of the soliton and a
more detailed account of the calculation will be published shortly.


\end{document}